\documentclass[cite,figures,doublecol]{epl2}
\usepackage{amsmath,amssymb}

\graphicspath{{./figures/}}

\title{Random RNA under tension}

\author{F. David\inst{1} \and C. Hagendorf\inst{1,2}
\and K.-J. Wiese\inst{2} \thanks{email:
\email{Francois.David@cea.fr, hagendor@lpt.ens.fr,
wiese@lpt.ens.fr}}}

\shortauthor{F. David \etal}

\institute{
\inst{1}{Service de Physique Th\'eorique\\CEA Saclay,
91191 Gif-sur-Yvette, France}\\\\
\inst{2}{CNRS-Laboratoire de Physique Th\'eorique de l'Ecole
Normale Sup\'erieure\\
24, rue Lhomond, 75231 Paris cedex 05, France}
}


\abstract{The L\"assig-Wiese (LW) field theory for the freezing
transition of random RNA secondary structures is generalized to the
situation of an external force. We find a second-order phase
transition at a critical applied force $f=f_{c}$. For $f<f_{c}$ forces
are irrelevant. For $f>f_{c}$, the extension $\cal L$ as a function of
pulling force $f$ scales as ${\cal L}(f) \sim
(f-f_{c})^{1/\gamma-1}$. The exponent $\gamma$ is calculated in an
$\epsilon$-expansion: At 1-loop order $\gamma = \epsilon/2=1/2$,
equivalent to the disorder-free case. 2-loop results yielding $\gamma
= 0.6$ are briefly mentioned. Using a locking argument, we speculate
that this result extends to the strong-disorder phase.}

\begin{document}
\maketitle
\section{Introduction}
RNA is a heteropolymer constructed from four different nucleotides
A, C, G and U located on a sugar-phosphate polymer backbone. In
solution, a single RNA strand bends back onto itself and folds into
a configuration of loops, stems and terminating bonds, due to
formation of Watson-Crick pairs A-U and C-G from bases located on
different parts of the strand. Together with environmental
conditions like temperature and ionic concentration, the
\textit{primary structure} (base sequence) determines the most
probable base-pairings, known as \textit{secondary structure}, which
then determines the most probable spatial conformation
(\textit{tertiary structure}) \cite{tinoco:99, higgs:00}. Unlike
protein folding, which exhibits a strong interdependence between
secondary and tertiary structure \cite{pande:00}, RNA folding may be
studied at the level of secondary structures due to a clear
separation of energy scales.

Since the pioneering work of Bundschuh and Hwa \cite{bundschuh},
several authors have studied the statistical physics of RNA
secondary structures for random sequences
\cite{paganini:00,krzakala:02,hui:06,monthus:06}. It is commonly
believed that these systems undergo a freezing transition upon
lowering the temperature. Based on a replica approach, L\"assig and
Wiese \cite{wiese:06}, and David and Wiese \cite{david:06} have
recently developed a systematic field-theory formulation for this
phase transition in terms of interacting random walks (RW). The
critical exponents characterizing pairing statistics and replica
overlap were computed within a 2-loop renormalization analysis, and
found to be remarkably close to numerical simulations
\cite{bundschuh,mueller:03}.

\begin{figure}
  \begin{center}
      \begin{tabular}{cc}
      \includegraphics[scale=0.8]{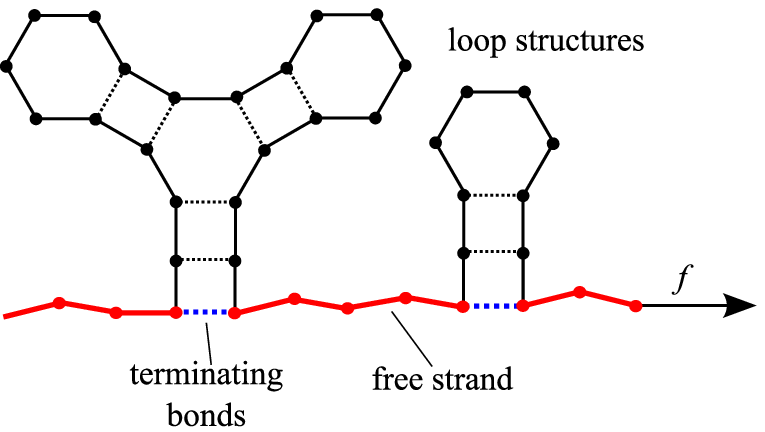} &(a)\\
      {}&{}\\
      \includegraphics[scale=0.8]{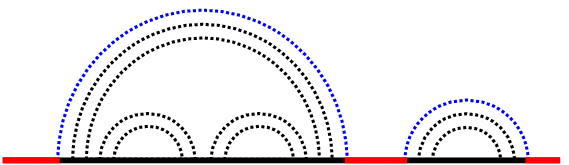} &(b)\\
      {}&{}\\
      \includegraphics[scale=0.8]{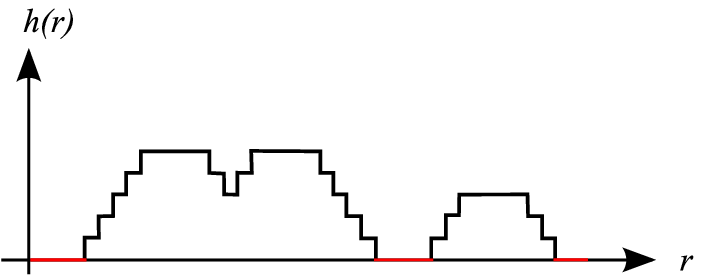} &(c)
      \end{tabular}
  \end{center}
  \caption{(a) Illustration of the planar structure of RNA
  under tension, (b) rainbow diagram and (c) corresponding
  height relief.}
  \label{fig:structures}
\end{figure}

An interesting way to probe RNA chains is to study its behavior
under an external pulling force (see fig.~\ref{fig:structures}a for
an illustration). Recently, force-extension curves have been
measured, by attaching beads to the RNA-molecule and pulling on it,
using an optical trap \cite{liphardt:01,onoa:03}. For homopolymers,
the competition between structure formation and denaturation of the
RNA strand leads to a second-order phase transition at a critical
force $f=f_c$ \cite{mueller:03}. For $f<f_c$ the strand is still in
a collapsed phase, while for $f>f_c$ it is in a extended ``necklace
phase" with a macroscopic extension (end-to-end distance). While
quite some literature exists about the subject (see e.g.
\cite{gerland:01,mueller:02b}) there is of today no theory to
compute the characteristic exponent $\gamma$ of the force-extension
characteristics for disordered RNA strands at the transition. In
this letter we fill this gap. We propose an extension of the RW
field theory for random RNA by including an external pulling force
$f$. We show that within our theory the force is renormalized by the
quenched disorder, hence the exponent $\gamma$ is modified with
respect to its mean-field value $\gamma_0=1/2$. Conversely, we argue
that the disorder coupling is not renormalized by the force at any
order in perturbation theory. We use perturbative renormalization of
the new theory to compute $\gamma$ to 1-loop order. Finally, we
comment on 2-loop results.
\section{The model}
We consider an RNA strand with $L$ bases labeled with indices
$s=1,\dots,L$. Similarly, we use the index $s$ to label a backbone
segment between adjacent bases $s$ and $s+1$. A secondary structure
$S$ is a set of base pairs $(s,t),\, s < t$. We only retain
so-called \textit{planar} structures $S$: any two different base
pairs $(s,t)$, $(s',t')$ are either independent $s < t < s' <t'$ or
nested $s < s' < t' < t$. $S$ can be represented by a diagram of
arches (fig. \ref{fig:structures}b). Planarity implies that the
arches do not cross each other. We introduce the \textit{contact
operator} $\Phi$ defined by $\Phi(s,t) = 1$ if $(s,t)\in S$, and $0$
otherwise \cite{wiese:06}. Moreover we define the \textit{height
field} $h$ on the segment $r$ by $h(r) := \sum_{s\le r}\sum_{t> r}^L
\Phi(s,t)$, counting the number of arches over $r$. This leads to an
identification of each open planar secondary structure $S$ with a
height function subject to boundary conditions $h(0)=h(L)=0$. A
segment between the bases $r$ and $r+1$ belongs to the free part of
the structure if and only if $h(r)=0$ (fig. \ref{fig:structures}c).

In order to develop a statistical mechanics model, we have to assign
to each structure $S$ an energy $E[S]$. We assume that it may be
written as a sum of the contributions from the formation of base
pairs $E_{\mathrm{pair}}[S]$ and from the external
force $E_{\mathrm{force}}[S]$. Bond formation between the bases at
$s$ and $t$ involves a pairing energy $\eta(s,t)$ which in general
depends on the nature of the pairing partners. We sum over all
pairing energies of base pairs in $S$  and obtain
$E_{\mathrm{pair}}[S] = \sum_{s<t}\eta(s,t)\Phi(s,t)$
\cite{bundschuh}. The energy due to the  external force,
$E_{\mathrm{force}}[S]$, depends  on the spatial configuration of
the free part of the strand and its elasticity. We assume that every
free backbone segment aligns with the force, hence the  energy is
proportional to the force $f$ times the number of monomers in the
free strand \cite{krzakala:02}. Thus, we neglect any elasticity and
entropic effects for the free segments and for the bonds which
terminate loop structures (figure \ref{fig:structures}a). By analogy
with the contact operator $\Phi(s,t)$, we introduce a free-strand
operator $\Delta(r)$  such that $\Delta(r)=1$ if $h(r)=0$, and $0$
otherwise. This allows to write
$E_{\mathrm{force}}[S]=-f\sum_r\Delta(r)\ .$

Having defined the energy of a given secondary structure, we proceed
to study the partition function
\begin{equation}
  Z = \sum_{S\in\mathcal{S}(L)}\exp\left[-\sum_{s<t}
  \eta(s,t)\Phi(s,t)+f\sum_s \Delta(s)\right]\label{eqn:partition}
\end{equation}
where $\mathcal{S}(L)$ denotes the set of all possible planar
secondary structures with $L$ bases. Before considering random RNA
chains, we briefly review the properties of the partition function
in the case of uniform pairing energies $\eta(s,t)=\eta_0$ (that we
may take $=0$). For $f=0$ one deduces from the height picture that
the problem is equivalent to the statistics of a RW on the positive
real axis $h\geq 0$. The partition function is $Z_L \propto
L^{-\rho_0}$ with $\rho_0 = 3/2$ the characteristic exponent of
first return. This leads to a pairing probability for the base pair
$(s,t)$ scaling like $p(s,t) =\langle \Phi(s,t)\rangle \propto
[|t-s|(L-|t-s|)]^{-\rho_0}$. Switching on the force $f>0$ amounts to
adding an attractive short-range potential at the origin
$h=0$. This is a well-known problem of statistical mechanics. For
instance it describes surface wetting transitions in $1+1$ dimensions
(see e.g.\ \cite{fisher:89}). For forces $f$ larger than a critical
force $f_c$, the RW is bound to the origin $h=0$ whereas for $f<f_c$
it is unbound (i.e.\ free to wander far away from $h=0$).

In fact, the $f=0$ problem can be mapped onto a free RW
$\vect{r}(s)$ in $d=2\rho_0=3$ dimensions; the height field is the
modulus $h(s) = |\vect{r}(s)|$. The inclusion of the short-ranged
attraction in the case $f>0$ corresponds to a short-ranged
attractive potential at $\vect{r}=0$. In the continuum limit, the
action of this model reads
\begin{equation}
  S_{3D}[\vect{r}(s)]=\int_0^L\upd s\,\left(
  \frac{1}{4}\left[\dot{\vect{r}}(s)
  \right]^2-f\delta^3(\vect{r}(s))\right)\ ,
  \label{eqn:height}
\end{equation}
and describes the pinning of a RW by an attractive impurity at the
origin.

We now exploit this analogy to extend our analysis to random RNA
structures. Numerical simulations \cite{bundschuh} suggest to model
sequence disorder by independent Gaussian random binding energies
$\eta(s,t)$,
\begin{equation}
\overline{\eta(s,t)}=\eta_0\ ,\ \
\overline{\eta(s,t)\eta(u,v)}-\eta_0^2 = \sigma^2 \delta(s-u)
\delta(t-v)
\end{equation}
Following \cite{wiese:06, david:06}, we construct a field theory in
the continuum limit $L\rightarrow \infty$. We  perform a
perturbative expansion in the disorder amplitude $g=\sigma^2>0$, and
the force strength $f$. To model disorder, we use the replica trick.
Each replica $\alpha=1,\dots,n$ is represented by a RW
$\vect{r}_{\alpha}(s)$ in an embedding space $\mathbb{R}^{d}$ (with
dimension $d=2\rho_0=3$). In fact, the explicit form of the pairing
probability for uniform RNA suggests to consider closed RWs
$\vect{r}_{\alpha}(s)$. Nevertheless, we can (and shall) use open
RWs because they have been proven to lie in the same universality
class and considerably simplify the calculations \cite{david:06}.
Within the RW representation, the contact operator reads
$\Phi_{\alpha}(s,t) = \delta^d
(\vect{r}_{\alpha}(s)-\vect{r}_{\alpha}(t))$. The average over the
disorder $\eta$ generates an attractive interaction between the
replicas. It is described by the \textit{overlap operator}
$\Psi_{\alpha\beta}(s,t) = \Phi_{\alpha}(s,t)\Phi_{\beta} (s,t)$
(counting the common arches of the replicas $\alpha$ and $\beta$ in
the original picture). By analogy with (\ref{eqn:height}) we
represent the operator $\Delta$ as
$\Delta_{\alpha}(s):=\delta^d(\vect{r}_{\alpha}(s))$. The resulting
action in the RW picture is
\begin{multline}
  S[\{\vect{r}_\alpha\}] = \sum_{\alpha}\int\frac{1}{4}
  \left[\dot{\vect{r}}_{\alpha}(s)\right]^2
  \\-g\sum_{\alpha<\beta}\iint_{s<t}
  \Psi_{\alpha\beta}(s,t)-f\sum_{\alpha=1}^n\int
  \Delta_{\alpha}(s)\ ,
\end{multline}
and generalizes the model of \cite{david:06} (where $f=0$). Before
using perturbation theory, we generalize the model to dimensions
$2\le d\le 3$ \cite{wiese:06,david:06}. Setting $\epsilon = d-2$ we
find the canonical scaling dimensions $\text{dim}\, g = \epsilon$
and $\text{dim}\, f =\epsilon/2$. The original theory corresponds to
$\epsilon=1$. The generalized model is renormalizable at
$\epsilon=0$ as the $f=0$ model \cite{DavidWiese2bePub}.
\section{Perturbation theory}
\begin{figure}[h]
  \begin{center}
    \begin{tabular}{cc}
      \includegraphics[height=1.cm]{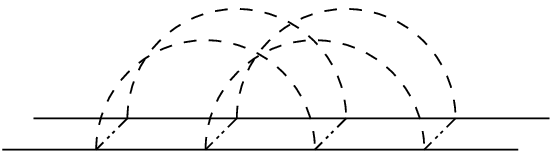}&
      \includegraphics[height=1.5cm]{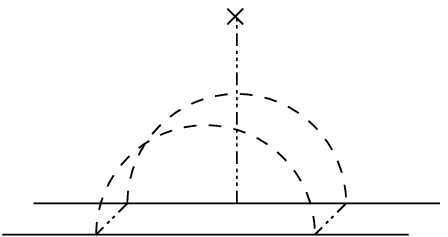}\\
      (a) & (b)
    \end{tabular}
  \end{center}
  \caption{Non-planar diagrams.}
  \label{fig:nonplanar}
\end{figure}

We represent the perturbative expansion of $Z$ in $f,\,g$ in terms
of Feynman diagrams. The $g$-vertex $\Psi_{\alpha\beta}$ (disorder
interaction vertex) is denoted by a double arch between a pair of
replicas \cite{wiese:06,david:06} and the $f$-vertex $\Delta_\alpha$
(force interaction at $r=0$) is depicted by a force insertion on a
single replica:
\begin{equation}
\Psi_{\alpha\beta}\ =\
\includegraphics[height=1.cm]{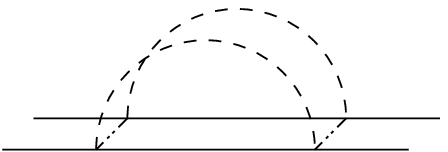}
\qquad\Delta_{\alpha}\ =\
\includegraphics[height =1.cm]{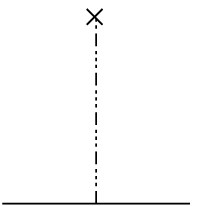}
\label{Vertices}
\end{equation}
Non-planar diagrams involving the disorder interaction, see fig.\
\ref{fig:nonplanar}(a), may be eliminated by introducing $n\times N$
pairs of auxiliary fields $\gamma$ and $\tilde\gamma$ in the action
and by taking the large-$N$ limit \cite{david:06}. Since the force
term only acts on the free part of the RNA strand, also diagrams of
the type given on fig.\ \ref{fig:nonplanar}(b) must be excluded.
This is achieved by a similar ``planarity constraint'' that can be
implemented using the same auxiliary fields.

We now consider the  partition function for a single (bundle of
replica) RW with fixed end-points, or rather its Fourier transform
$Z(\{\vect{q}_{\alpha}'\},\{\vect{q}_{\alpha}''\};f,g)$, defined as
\begin{multline}
Z(\{\vect{q}_{\alpha}',\vect{q}_{\alpha}''\};f,g) \\=
\int\mathcal{D}[\{\vect r_\alpha\}]\,\mathrm{e}^{-S[\{\vect
r_\alpha\}]}\, \prod_{\alpha}V_{\vect q'_{\alpha}}^\alpha(0)
V_{\vect q''_{\alpha}}^\alpha(L)
\end{multline}
where the vertex operator $V_{\vect q}^\alpha(s)=\exp\left(i\vect
q\vect r_\alpha(s)\right)$ injects incoming external momenta
$\{\vect{q}_{\alpha}'\},\{\vect{q}_{\alpha}''\}$ to each end-point
of the replica $\alpha$ of the RW. The (regularized) dimensionless
integration measure is given by $\mathcal{D}[\vect{r}_{\alpha}] =
\prod_i \upd^d \vect{r}_{\alpha,i}/(4\pi a^2/m)^{d/2}$ where $a$ is
an ultraviolet cut-off and $m = 1/2$ the mass of the Brownian
particle associated with the RW. We further simplify the model by
setting $\vect{q}_{\alpha}'\equiv
\vect{q}',\,\vect{q}_{\alpha}''\equiv\vect{q}''$ without loss of
generality.

The perturbative expansion in $g$ and $f$ leads to a systematic
diagrammatic representation of $Z$. The diagrams can be classified
according to the number of replicas with force insertions. The
absence of force insertions on the replica $\alpha$ leads to a
momentum conservation
$\delta^{d}(\vect{q}'_{\alpha}+\vect{q}''_{\alpha})$ (translation
invariance $\vect r_\alpha(s)\to\vect r _\alpha(s)+\vect r$). Thus,
the set of all possible diagrams is classified according to the
number $k$ of its external momentum conservations: $k=n$
conservations correspond to a diagram of the force-free theory,
$k=n-1$ conservations to a single replica subject to force
insertions, etc. We group all diagrams with $k$ conservations into a
restricted partition function $\Xi_{n-k}(\vect{q}',\vect{q}'';f,g)$,
so that formally $Z(\vect{q}',\vect{q}'';f,g)=\sum_k
\delta(\vect{q}'+\vect{q}'')^k\Xi_{n-k}(\vect{q}',\vect{q}'';f,g)$.
$\Xi_0$ describes the force-free theory, $\Xi_1$ the sector where
the attractive short-range potential only acts upon a single
replica, $\Xi_2$ the sector where it acts upon two replicas, etc. In
the following we focus on $\Xi_1(\vect q,-\vect q)$, since this
simplifies the calculations. Perturbative expansion up to order two
in $f$ or $g$ yields
\begin{align}
\label{eqn:dev}
&\Xi_1 (\vect q,-\vect q;f,g) =
\nonumber
\\
&\ f\,n\,\includegraphics[height=1cm]{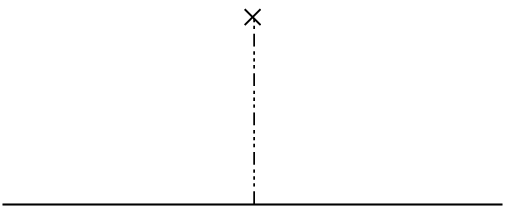}\,+\,
f^2\,n\,\includegraphics[height=1cm]{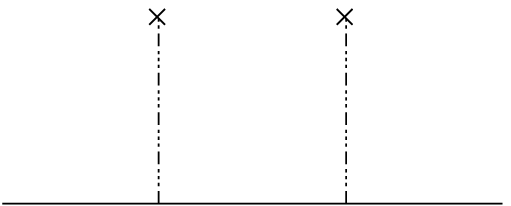} \nonumber
\\
&+
\,fg\Bigg[n(n-1)\,\bigl(\
\includegraphics[height=1cm]{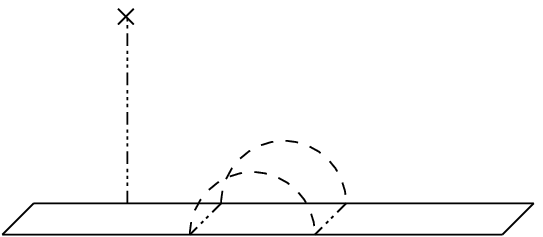}
+
\includegraphics[height=1cm]{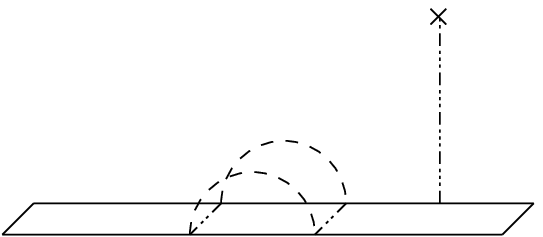}
\bigr)\nonumber\\
&\qquad~~~+ \frac{n(n-1)(n-2)}{2}\
\includegraphics[height=1cm]{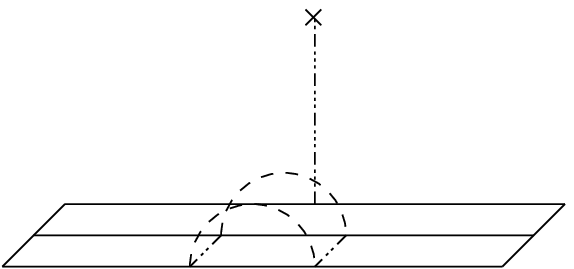} \Bigg] \,+\,\cdots
\end{align}
There are four topologically different Feynman diagrams.
The first contribution is
\begin{align}
\includegraphics[height=1cm]{figure5}&=\underset{0<u<L}{\int}
\hskip -0.5em \langle {\Delta_1(u)\prod_{\alpha=1}^n V_{\vect
q}^\alpha(0)V_{-\vect q}^\alpha(L)}\rangle_0 \nonumber\\
&=\ L\,e^{-{\vect q}^2Ln} , \label{eqn:firstdiagr}
\end{align}
where $\langle\quad\rangle_0$ denotes the average w.r.t.\ the free RW
action $S_0[\vect r]$ ($f=g=0$), with a proper subtraction of the
$(n-1)$ translational zero-modes. The second diagram is UV divergent
at $\epsilon=0$. Isolation of the corresponding pole yields
\begin{align}
\label{} &\includegraphics[height=1cm]{figure6} \nonumber\\
&=\underset{0<u<v<L}{\iint}\hskip -0.5em \langle
{\Delta_1(u)\Delta_1(v) \prod_{\alpha=1}^n V_{\vect
q}^\alpha(0)V_{-\vect q}^\alpha(L)}\rangle_0
\nonumber\\
&\ =\ \underset{0<u<v<L}{\iint}e^{-\vect
q^2(L-v+u)}|v-u|^{-d/2}e^{-\vect q^2L\,(n-1)} \nonumber\\
& \mathop{=}_{\epsilon \to 0}\ -\frac{L^2}{\epsilon}\, e^{-{\vect
q}^2Ln}+O(1)
\end{align}
The divergence comes from the short-distance behavior of the product
of two $\Delta$ operators $\Delta(u)\Delta(v)\mathop{=}\limits_{u\to
v}|u-v|^{-d/2}\Delta(v)+\cdots$. The third contribution is also UV
divergent,
\begin{align}
&\includegraphics[height=1cm]{figure7}=
\includegraphics[height=1cm]{figure8}\nonumber\\
&=\underset{0<u<v<w<L}{\iiint}\langle {\Delta_1(u)\Psi_{12}(v,w)
\prod_{\alpha=1}^n V_{\vect q}^\alpha(0)V_{-\vect
q}^\alpha(L)}\rangle_0
\nonumber\\
&= \underset{0<u<v<w<L}{\iiint}e^{-2\vect
q^2(L-w+v)}|w-v|^{-d}e^{-\vect q^2 L\,(n-2)}
\nonumber \\
&\mathop{=}_{\epsilon\to 0}\ -\frac{L}{\epsilon}\, (L\vect
q^2-1)\,e^{-{\vect q}^2Ln}+O(1)\ ,
\end{align}
as well as the fourth
\begin{align}
\label{eqn:lastdiagr}
&\includegraphics[height=1cm]{figure9}\nonumber\\
&= \underset{\substack{0<u<L\\0<v<w<L}}{\iiint} \langle
{\Delta_1(u)\Psi_{23}(v,w) \prod_{\alpha=1}^n V_{\vect
q}^\alpha(0)V_{-\vect q}^\alpha(L)}\rangle_0
\nonumber\\
&=\ \underset{\substack{0<u<L\\0<v<w<L}}{\iiint}e^{-2\vect
q^2(L-w+v)}|w-v|^{-d}e^{-\vect q^2L\,(n-2)}\nonumber\\
&\mathop{=}_{\epsilon\to 0} -\frac{L}{\epsilon}\, (2L\vect
q^2-1)\,e^{-{\vect q}^2Ln}+O(1)\ .
\end{align}

\section{Renormalization} 
\begin{figure}
  \begin{center}
    \includegraphics{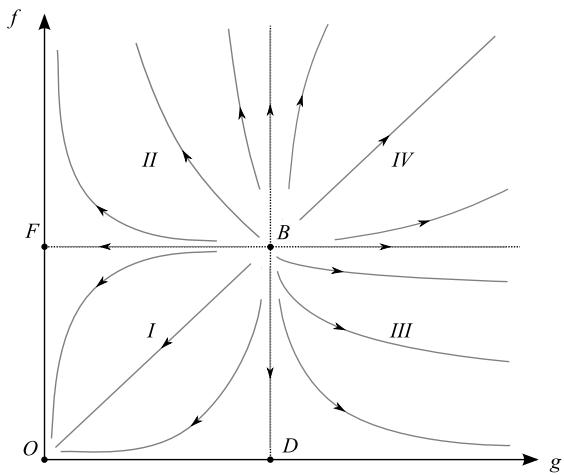}
  \end{center}
  \caption{Schematic view of renormalization group flow in the $(g,f)$
  plane. The critical lines distinguish the molten phase (I), a
  glass phase (II), (III) the denatured homopolymer phase. A new
  phase (IV) corresponds to RNA strands with isolated frozen
  structures separated by free parts.}
  \label{fig:flow}
\end{figure}

We remove the UV divergences in the
expansion (\ref{eqn:dev}) by formally taking $\rho_0$ as
analytical regularization parameter. In order to eliminate the
simple poles in $\epsilon = 2\rho_0 -2$ at $\epsilon=0$, we define
the renormalized theory through the renormalized action
\begin{multline}
S_R = \sum_{\alpha}\int_s
\frac{\mathbb{Z}}{4}\dot{\vect{r}}^2_{\alpha}
-g_R\mu^{-\epsilon}\mathbb{Z}_g
\sum_{\alpha<\beta}\underset{0<s<t<L}{\iint}\Psi_{\alpha\beta}(s,t)\\
- f_R\mu^{-\epsilon/2}\mathbb{Z}_f\int_s \Delta_{\alpha}(s) + 2
n\mathbb{Z}_1\ .
\end{multline}
Here $\mathbb{Z}$, $\mathbb{Z}_g$ and $\mathbb{Z}_f$ denote the
wave-function, the coupling constant and the force counterterms
respectively. $\mathbb{Z}_1$ accounts for boundary effects since we
deal with \textit{open} RWs \cite{david:06}. The coefficients of
their development in $f_R,\,g_R$ contain the leading poles in
$1/\epsilon$. Furthermore, we have introduced the renormalization
mass scale $\mu\sim 1/L$. From dimensional analysis we deduce the
relations between renormalized and bare $\vect{r}$, $\vect{q}$, $g$
and $f$: For the field $\vect{r} = \mathbb{Z}^{1/2} \vect{r}_R$,
$\vect{q} = \mathbb{Z}^{-1/2} \vect{q}_R$, for the couplings $g =
g_R\mu^{-\epsilon}\mathbb{Z}_g\mathbb{Z}^{-d}$ and $f =
f_R\mu^{-\epsilon/2}\mathbb{Z}_f\mathbb{Z}^{-d/2}$. The renormalized
partition functions are related to the bare ones via
\begin{equation}
  \Xi_k^R(\vect{q}'_R,\vect{q}''_R,f_R,g_R)=\mathbb{Z}^{-kd/2}
  e^{-2n\mathbb{Z}_1}\Xi_k(\vect{q}',\vect{q}'',f,g)\ .
\end{equation}
The prefactor on the r.h.s.\ takes into account the $(n-k)$ zero
modes, boundary effects and the change of normalization in the
integration measure. Consistency of the theory requires cancelation
of all divergences upon renormalization for each individual $\Xi_k$.
In particular, renormalization of the force-free terms $\Xi_0$ has
been performed previously \cite{david:06} and yields the counter
terms $\mathbb{Z}=1+g_R(n-1)/\epsilon$,
$\mathbb{Z}_g=1+g_R(7-4n)/\epsilon$ and
$\mathbb{Z}_1=1+3g(n-1)/4\epsilon$. They do not depend on $f$ at any
order since they correspond to ``bulk'' divergences for the random
walk in interaction with the potential at the origin (``boundary''
term). For the counter term $\mathbb{Z}_f$ we need $\Xi_{1}$, whose
Feynman diagrams were computed in eqns.\ (\ref{eqn:firstdiagr})-(\ref{eqn:lastdiagr}):
\begin{multline}
  \Xi_{1}
  = nLf e^{-(\vect{q}')^2Ln}\\\times\left[1+\frac{g (n-1)}{2\epsilon}
  \left(\frac{n+2}{2}-2n(\vect{q}')^2L\right)
  -\frac{2f}{\epsilon}\right].
\end{multline}
We absorb the poles by the counter terms of the force-free theory
and $\mathbb{Z}_f = 1 + 2f/\epsilon$. This expression depends
neither on the number of replicas $n$, nor on $g$. Thus at first
order there is no coupling between force and disorder. A
straightforward calculation gives the RG $\beta$-functions
\begin{align}
\beta_f(g_{R},f_{R}) &=   -\left.\frac{\upd f_R}{\upd \log
\mu}\right|_{f}=-\frac{\epsilon}{2}f_{R}+f_{R}^2,\\
\beta_g(g_{R},f_{R}) &= -\left.\frac{\upd g_R}{\upd \log
\mu}\right|_{g}=-\epsilon g_{R} + (5-2n)g_{R}^2
\end{align}
In the physical case of random RNA, $n=0$, they yield the RG flow
depicted in fig. \ref{fig:flow} with four fixed points. The
attractive Gaussian fixed point $O = (0,0)$ describes the molten
phase of \cite{bundschuh}. $D = (\epsilon/5,0)$ is the fixed point
of the glass transition found in \cite{wiese:06, david:06}. We
identify $F = (0,\epsilon/2)$ with the denaturation transition of a
homopolymer (wetting in $1+1$ dimensions). More interesting, a new
bi-critical UV unstable fixed point $B =(\epsilon/2,\epsilon/5)$
emerges from our one-loop analysis. It leads to four phases
separated by the critical lines $f \equiv \epsilon/2$ and $g \equiv
\epsilon/5$. In particular, a new phase at both high force and
strong disorder emerges. Physically, it corresponds to isolated
frozen branched structures separated by free parts of the strand.

The extension ${\cal L}$ as a function of the force $f$ is given by
the anomalous dimension of this force. The scaling ansatz reads
\begin{equation}
{\cal L}(f) = L^{\alpha} \mathcal{F}\left([f-f_{c}]
L^{\gamma}\right)\ ,
\end{equation}
where $\mathcal{F}(z)$ is a scaling  function. At the bi-critical
fixed point, we find
\begin{equation}\label{scaling-function}
\gamma= -\dim [f-f_{c}] = \left.\frac{\partial \beta_f}{\partial
f_R}\right|_{f_c} = \frac{\epsilon}{2}\ .
\end{equation}
Since ${\cal L} =\left\langle\int_{0}^{L}\mathrm{d} x\,
{\Delta(x)}\right\rangle$, the exponent $\alpha=\gamma$. Demanding
that for large systems the extension ${\cal L}$ become extensive,
i.e.\ $\sim L$, yields $\mathcal{F}(z)\sim z^{\frac 1 \gamma -1}$
for $z \gg 1$, so that for large RNA molecules $ {\cal L} (f) \sim L
(f-f_{c})^{\frac 1\gamma -1}$. Setting $\epsilon = 1$ yields the
exponent $\gamma =1/2$ which is the same result as for the
homopolymer denaturation transition, first discussed in
\cite{mueller:03}. This is consistent with the fact that the 1-loop
beta function for $f$ does not depend on $g$, so that $\gamma$ is
not changed by the quenched disorder. However, this result does not
hold at higher orders \cite{DavidWiese2bePub}.

\section{Conclusion}
To summarize, we have developed a field-theoretic description of
random RNA under application of an external force, by extending the
force-free field theory. It permits to study the second-order
denaturation transition. At 1-loop order, the RG flow functions for
force $f$ and disorder strength $g$ decouple so that the
denaturation is not influenced by disorder. We have computed the
critical exponent $\gamma = 1/2$ for the force-extension
characteristic at the transition which agrees with previously
considered homopolymer models.

We have extended our calculations to second order in perturbation
theory \cite{DavidWiese2bePub}. The renormalization group flow of
$f$ then depends on the disorder strength $g$. The procedure yields
a scaling exponent $\gamma = 0.6$ for the force-extension
characteristic, resulting in ${\cal L} (f) \sim L (f-f_{c})^{\frac
1\gamma -1}\approx L (f-f_{c})^{2/3} $.

We invoke the locking hypothesis \cite{wiese:06} in order to
conjecture the critical exponents in the glass phase. The argument
relies on the exact inequality $dim[\Psi]\ge dim[\Phi]$ for the
dimensions of contact and overlap operators. For $\epsilon=1$
consistency of the theory requires that the inequality is satisfied.
Physically this means that different replicas follow the same
minimal-energy path already at the transition. This is quite
unusual. Normally one expects several minimal-energy paths to exist
at the transition, and the strong-coupling phase to be characterized
by a collapse of these distinct paths into a single one, resulting
into different physical exponents. In a situation where at the
transition only a single minimal-energy path exists, we cannot have
a collapse of several paths, and the accompanying change of critical
exponents. In the absence of a different mechanism, the exponents
will not change for a small increase in disorder, and by
renormalization arguments in the whole strong-coupling phase. We
conjecture that this is what happens for the RNA freezing
transition. It is then tempting to suppose this hypothesis to hold
even in the presence of an external force, i.e.\ on the critical
line beyond the bi-critical fixed point. This assumption leads us to
extend our prediction $\alpha=\gamma=0.6$ to the glass phase.
Indeed, this prediction proves to be in reasonable agreement with
numerical simulations of M\"uller \etal
\cite{mueller:03,mueller:06}.

\acknowledgements This work is supported by the EU ENRAGE network
(MRTN-CT-2004-005616) and the Agence Nationale de la Recherche
(ANR-05-BLAN-0029-01 \& ANR-05-BLAN-0099-01). We thank the KITP (NSF
PHY99-07949) where part of this work was done. We also thank Markus
M\"uller for stimulating discussions.

\end{document}